\documentclass{vgtc}                          % final (conference style)
\ifpdf%                                % if we use pdflatex
  \pdfoutput=1\relax                   % create PDFs from pdfLaTeX
  \usepackage{graphicx}                % allow us to embed graphics files
  \DeclareGraphicsExtensions{.pdf,.png,.jpg,.jpeg} % for pdflatex we expect .pdf, .png, or .jpg files
\else%                                 % else we use pure latex
  \usepackage{graphicx}                % allow us to embed graphics files
  \DeclareGraphicsExtensions{.eps}     % for pure latex we expect eps files
\fi%

\graphicspath{{figures/}{pictures/}{images/}{./}} % where to search for the images

\usepackage{microtype}                 % use micro-typography (slightly more compact, better to read)
\PassOptionsToPackage{warn}{textcomp}  % to address font issues with \textrightarrow
\usepackage{textcomp}                  % use better special symbols
\usepackage{mathptmx}                  % use matching math font
\usepackage{times}                     % we use Times as the main font
\usepackage{cite}                      % needed to automatically sort the references
\usepackage{tabu}                      % only used for the table example
\usepackage{booktabs}                  % only used for the table example
\usepackage{color}
\newcommand{\ignore}[1]{}
\newcommand{\notes}[1]{}%\textcolor{blue}{#1}}

\newcommand{\task}[1]{ \noindent \\ {\it \textbf{Downstream Task: #1}}}

\newcommand{\usecase}[1]{{\it #1}}

\newcommand{\taskdetails}[1]{ \noindent \\ #1}
 
\newcommand{\fixtitle}{~~~~}

\usepackage{xspace}

\newcommand{\ho}{{\em Human only}\@\xspace}
\newcommand{\mo}{{\em Machine only}\@\xspace}
\newcommand{\hm}{{\em Human + Machine}\@\xspace}
\newcommand{\hme}{{\em Human + Machine + Explanation}\@\xspace}

\onlineid{0}

\vgtccategory{Research}

\vgtcinsertpkg

\title{Measure Utility, Gain Trust: Practical Advice for XAI Researchers}

\author{Brittany Davis\thanks{e-mail: brittany.f.davis@wsu.edu}\\ %
\scriptsize School of Engineering and Applied Sciences \\Washington State University %
\and Maria Glenski\thanks{e-mail: maria.glenski@pnnl.gov}\\ %
\scriptsize Data Sciences and Analytics Group \\ Pacific Northwest National Laboratory %
\\
\and \fixtitle William Sealy\thanks{e-mail: william.sealy@gatech.edu}\\ %
\scriptsize \fixtitle Cognitive Engineering Center \\ \fixtitle Georgia Institute of Technology
\and \fixtitle Dustin Arendt\thanks{e-mail: dustin.arendt@pnnl.gov}\\ %
\scriptsize \fixtitle Visual Analytics Group \\ \fixtitle Pacific Northwest National Laboratory}

\abstract{%
Research into the explanation of machine learning models, i.e., explainable AI (XAI), has seen a commensurate exponential growth alongside deep artificial neural networks throughout the past decade. For historical reasons, explanation and trust have been intertwined. However, the focus on trust is too narrow, and has led the research community astray from tried and true empirical methods that produced more defensible scientific knowledge about people and explanations. To address this, we contribute a practical path forward for researchers in the XAI field. We recommend researchers focus on the utility of machine learning explanations instead of trust. We outline five broad use cases where explanations are useful and, for each, we describe pseudo-experiments that rely on objective empirical measurements and falsifiable hypotheses. We believe that this experimental rigor is necessary to contribute to scientific knowledge in the field of XAI.} % end of abstract

\CCScatlist{
  \CCScatTwelve{Human-centered computing}{Human computer interaction (HCI)}{HCI design and evaluation methods}{User studies};
  \CCScatTwelve{Computing methodologies}{Machine learning}{Machine learning approaches}{Neural networks}
}

\begin{document}

%% The ``\maketitle'' command must be the first command after the
%% ``\begin{document}'' command. It prepares and prints the title block.

%% the only exception to this rule is the \firstsection command
\firstsection{Position: Measure Utility, Gain Trust}
\maketitle

 Many AI, HCI, and Visualization researchers may have assumed the purpose of a machine learning explanation is to enhance, increase, or calibrate users' trust in the model~\cite{chatzimparmpas2020state}.
 In contrast with this viewpoint and the large body of existing research, we argue that trust is an insufficient metric for evaluating explanations (or models). We recommend that researchers objectively measure the utility of the explanation instead of subjectively measuring trust. Trust should manifest through experience~---~as users use a system containing a model and an explanation, their trust in the system will grow if that system is reliable and provides a benefit. We also suspect that designing explanations to optimize trust may short-circuit the natural process of trust-building and also mislead users. This could occur if the design of the explanation obfuscates the actual utility or capability of the system. Optimizing for trust could be considered a form of ``metric hacking'', leading to artificially inflated levels of trust compared to what would build naturally through experience.
 
 If users' increased trust in the model is not necessarily the most appropriate way to measure the ``goodness'' of an explanation, then what is? Do explanations have an intrinsic value? We believe the answer to the latter question is simply no. Unlike machine learning models, which are built directly upon ground truth data, a ground truth ``correct'' explanation of a machine learning model is not generally available. Thus, we are not likely to find a way to measure the error between a given explanation and the correct one, except in very specific use cases, e.g., image captioning. Much like a data visualization, the purpose of an explanation is to communicate useful information to a human~---~the visualization community has no methods for directly measuring the correctness of a visualization. We do however have many methods to measure the utility of visualization, which all require considering what tasks users perform with that visualization. In fact, many explanations {\em are} visualisations~\cite{hohman2018visual}, so the full range of techniques the visualization community has employed for evaluating visualizations remain relevant for evaluating machine learning explanations.
 
 There are many use cases where explanations are helpful as part of a larger workflow. Hohman et al.~\cite{hohman2018visual} and Mohseni et al.~\cite{mohseni2018survey} both describe in depth the various users and uses of explanations. For the purposes of our argument in this paper, we focus on the following use cases: debugging, validating, and selecting a model; understanding a model; teaming with a model; and challenging a model. Each of these use cases allow us to imagine ``downstream tasks'' where we can quantitatively measure the users' performance with or without the explanation. Resulting differences in performance provide indirect but compelling evidence of the utility of the explanation.
There are many potential ways of measuring ``explanation goodness'', which interested readers can find surveyed by Mohseni et al.~\cite{mohseni2018survey}. Hoffman et al.~\cite{hoffman2018metrics} also discuss several methods for evaluating explanations including those that go beyond trust.

We have taken the position that trust is a flawed metric for measuring the ``goodness'' of a machine learning explanation. The next section of this paper will provide historical context for this argument. We found that trust is a metric advocated by people who want the models they build to be adopted (thus, a good explanation is one that builds trust and increases the likelihood of adoption). The final section of this paper provides practical guidance for how to measure the utility of an explanation across the use cases mentioned above.

\section{The Past: historical reasons for the coupling of trust and explanations}

% todo: summary of sections main points (Britt) (read once more)
%   

While the success of deep neural networks in various machine learning tasks is a recent phenomenon, machine learning as a general practice has existed for much longer. Almost since the beginning, developers of machine learning algorithms have sought to increase the acceptance and uptake of those algorithms. To do so, researchers had to prove to others what they felt they knew to be true~---~that these algorithms could be trusted to work correctly and had benefit. Thus, explanations of classical machine learning were initially developed for the purpose of increasing users' trust. As the field changed over time, the coupling of explanations to trust-building became a habit more than best-practice. Today, explanations can and should be used more widely beyond increasing trust~---~measuring their effectiveness should account for this broader applicability.

\subsection{``Classical'' Machine Learning}

In 2000, AlexNet had not yet won an image classification competition and the field of artificial intelligence (AI) was focused on things like ``symbolic systems'', which were built on pieces of modular knowledge, gleaned from experts or large databases and stored into a searchable structure such as a decision tree. Herlocker et al.~\cite{herlockerExplainingCollaborativeFiltering2000} wrote that ``current recommender systems are black boxes, providing no transparency into the working of the recommendation. Explanations provide that transparency, exposing the reasoning and data behind a recommendation'' and presented an experiment that measured which explanations increased consumers’ acceptance of recommendations generated by an intelligent agent and which ``negatively contributing to the acceptance of the recommendation.'' These systems may appear to be mysterious to end users but they were just data structures to the researchers who built and deployed them. Researchers knew how to construct and debug these structures, because it was the same way that they constructed and debugged code. 

However, these researchers needed a way to get the skeptical public — including corporations — to accept these expert systems. As Bilgic and Mooney~\cite{bilgicExplainingRecommendationsSatisfaction2005} put it in 2005, ``in order for users to benefit, they must trust the system’s recommendations and accept them. A system’s ability to explain its recommendations in a way that makes its reasoning more transparent can contribute significantly to users’ acceptance of its suggestions.'' This was the apparent goal of trust with pre-neural-network-AI: to help purchasers and users to understand the limits of the system so that an inevitable wrong answer would not scare them away. By 2006, Pu and Chen~\cite{puTrustBuildingExplanation2006} referred to this pursuit as ``investigating design issues for trust-inducing interfaces,'' and in 2009, Haynes et al.~\cite{haynesDesignsExplainingIntelligent2009} observed that ``[a]s intelligent agents become more pervasive in our day-to-day computing environment, and as their role becomes more consequential with respect to human purposes, they will be increasingly called upon to communicate in a way that engenders trust''

%\subsection{Trust as a vehicle for adoption}%Trust in pursuit of Adoption}%

\subsection{Deep Learning and the Image Domain}%Influence of Deep Learning and the Image Domain}
In 2012, AlexNet achieved a top-5 error rate of 15.3\% in the ImageNet Large Scale Visual Recognition Challenge \cite{krizhevskyImageNetClassificationDeep2017} and by 2013 researchers were trying to peer inside the Convolutional Neural Networks which made up AlexNet. Why? Because, as Zeiler and Fergus~\cite{zeilerVisualizingUnderstandingConvolutional2013a} noted less than a year later, ``there is no clear understanding of why they perform so well, or how they might be improved.'' Or, as Holzinger et al.~\cite{holzingerCausabilityExplainabilityArtificial2019} put it,``Technically, the problem of explainability is as old as AI itself and classic AI represented comprehensible retraceable approaches. However, their weakness was in dealing with uncertainties of the real world. Through the introduction of probabilistic learning, applications became increasingly successful but increasingly opaque.'' Researchers were suddenly in the same position as those suspicious consumers: they weren't sure why or how neural networks worked, or what tweaks to their construction might improve or deteriorate performance. In response, researchers began trying to find ways to prove to themselves that these new `expert’ systems were trustworthy.

From figuring out which layers were detecting edges versus texture \cite{simonyanDeepConvolutionalNetworks2014a}, to generating generalized images of what specific neuron clusters were capturing \cite{nguyenSynthesizingPreferredInputs2016}, or understanding what changes could produce wrong answers in adversarial attacks, researchers in AI had to go back to basics when neural networks exploded onto the scene. By 2016, Zeiler and Fergus’s~\cite{bachPixelWiseExplanationsNonLinear2015} method of pushing convolutions backwards in a network had inspired the development of Layer-wise Relevance Propagation. That same year, Ribiero et al.~\cite{ribeiro2016should} published their methodology, called LIME, for generating local or global counter-explanations for image classifiers. In 2017, a methodology called Grad-CAM was developed, which used gradients to highlight parts of the input to an image classifier which contributed most to the outcome \cite{selvarajuGradCAMVisualExplanations2017}. All of these methods were used by the research community to try and better understand the inner workings of deep neural networks. 

Most work on end-user-facing trust in this period was still analyzing trust in symbolic systems and other more established machine learning techniques. These experiments still largely aimed to increase user’s trust in the systems, by understanding what modulated user trust up or down. Researchers tested explanations of simplified models of bagged decision trees, and found that ``when soundness was very low, participants experienced more mental demand and lost trust in the explanations''~\cite{kuleszaTooMuchToo2013}. Another experiment tested explanations of a system which added a probabilistic Markov Decision Process to a task planning application based on a finite state machine, and found ``that transparency explanations can help to reduce the negative effects of trust loss \cite{nothdurftProbabilisticHumanComputerTrust2014}.'' Some experiments used the ``Wizard of Oz'' approach to test user trust in automated systems, where ``the behavior of the software is controlled by the researcher unbeknown to the participant \cite{bussoneRoleExplanationsTrust2015}.'' Another experiment conducted in 2016 used an Auto-Encoder developed in 2010 to find ``that perceived system ability was more important in determining trust amongst with-explanation participants and perceived transparency was a greater influence on the trust of participants who did not receive explanations \cite{hollidayUserTrustIntelligent2016}.'' Berkovsky et al.~\cite{berkovskyHowRecommendUser2017} ran an experiment ``to investigate the dependencies between various aspects of recommendation interfaces and user-system trust.''

\subsection{The Shift to \textit{Appropriate} Trust}
In August of 2016, DARPA announced its Explainable Artificial Intelligence program. The announcement brought into common use among researchers the phrase ``appropriate trust” and the idea of explainable AI among the general population \cite{gunningBroadAgencyAnnouncement2016}. Around the same time, the European Union (EU) had announced that the General Data Protection Regulation (GDPR)
would take effect in 2018 \cite{wolford}. This regulation specified that companies could no longer use systems to make certain decisions about consumers who lived in the EU, unless the company could explain the decision. These two events seem to have spurred researchers to tentatively turn their efforts at explaining neural networks outside of the research community, engaging end users and sometimes everyday people.

Researchers began to comb over the progress made on peering inside of neural networks and trying to find ways to use these tools to increase appropriate trust among people outside the AI community. For example, Schaefer et al.~\cite{schaeferCommunicatingIntentDevelop2017} found that ``by understanding the transparency elements that increase effective bi-directional communication [in human-computer teams], we can… engender appropriate trust and reliance in the system.'' 

In addition, much debate ensued over defining, justifying, and measuring ‘explainable’ and ‘trust’ in this new context. Jiang et al.~\cite{jiangTrustNotTrust2018} presented the concept of a trust score, to provide ``information about the relative positions of the data points, which may be lost in common approaches such as the model confidence when the model is trained using stochastic gradient descent.'' A plethora of taxonomies for trust measurements and explanations appeared. Adadi and Berrada~\cite{adadiPeekingBlackBoxSurvey2018} claimed explanations are needed for four reasons: to justify, to control, to improve, and to discover. Pallotta et al.~\cite{pallottaSmartHeatingSystems2008} argue that ``explanations need to be carefully crafted to fit with their desired aim,” and described a methodology which would increase user trust enough to prevent users from interfering with home heating systems. Holzinger et al.~\cite{holzingerCanWeTrust2018} argued that increasing trust in deep learning systems necessarily included mechanisms for users to change the system’s outcomes.  

Out of this debate emerged a growing consensus that the emphasis should be on the word `appropriate’ in the phrase ‘appropriate trust’. Yin et al.~\cite{yinUnderstandingEffectAccuracy2019} measured user’s trust in a model and found that both the actual capabilities of the model and the specific instances seen by the user influence a user’s trust in the model. That is to say, if a model is not accurate, and this is evident to users, they don’t trust it — and that’s a good thing. However, another experiment found that users can rely too heavily on a poor model, reporting that ``in 67.3\% of all cases, participants predicted that the system would be correct, whereas it was only correct in 42.9\% of the cases \cite{alqaraawiEvaluatingSaliencyMap2020}.''

%\textbf {Try to connect to our position here, something like...}
\vspace{0.5\baselineskip}

Neural networks and deep learning shifted the paradigm in such a way that engendering trust, even appropriate trust, is no longer sufficient. The idea of trust is already ambiguous, and applying it to deep neural networks when even those who build them do not understand the intricacies of the model logic or determinations creates too many layers of abstraction to produce meaningful science. Instead, researchers should re-evaluate the revelations of 2013, and be humbled by the fact that we still do not know how to debug or improve these networks with much certainty. To that end, creating falsifiable and provable hypotheses should replace the concept of increasing or calibrating trust. Although appropriate trust is a more objective measurement of trust, we ought to be measuring these systems with more relevant metrics that can be clearly measured, tested, and replicated.

\section{A Possible Future: Downstream Tasks and Falsifiable Hypotheses}
As previously argued, the utility of an explanation must be tied to its purpose~---~why it was created and the context in which it is intended to be used. Trust should not be a metric we maximize through design, but should be a benefit gained after a user interacts with a useful system over time. Thus we believe that studies evaluating explanations should not measure trust unless they are longitudinal, ``in the wild'', and consider the entire system. Instead, we argue that research studies that seek to measure the benefit of novel explanations should focus on utility over trust. We have identified five broad use cases where researchers could design experiments to measure the utility of explanations. 
\begin{enumerate}
\itemsep-0.1em
    \item {\it Model Debugging and Validation}: Is the model working as designed? Why is the model making mistakes?
    \item {\it Model Selection}: Potentially going beyond simple performance metrics like F-score, which model is best?
    \item {\it Mental Model and Model Understanding}: How does the model function or behave? Can I learn something interesting from the model?
    \item {\it Human Machine Teaming}: Can I do a task with the model better than on my own (and better than the model on its own)?
    \item {\it Model Feedback, Challenging, and Prescription}: When I am affected by a model's decision, how do I challenge that decision or correct the model when it's wrong about me? What should I change about me to get a better outcome in the future?
\end{enumerate}

\begin{figure}
    \centering
    \includegraphics[width=0.45\textwidth]{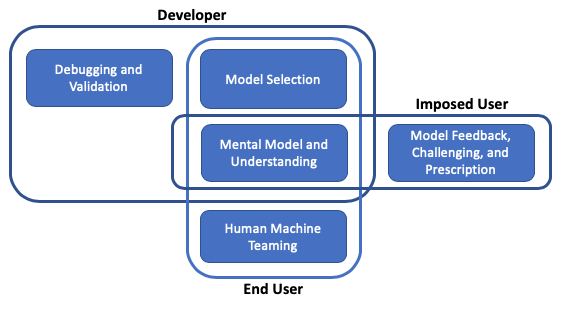}
    \caption{Venn diagram illustrating the overlap in interest across use cases for differing viewpoints -- from the \textit{developer} who builds the model, the \textit{end user} who selects and applies the model, or the \textit{imposed user} who is affected by the decisions or recommendations the model provides. }
    \label{fig:use_case_diagram}
\end{figure}

The utility of a model and associated explanations can be measured from several viewpoints. We focus on three: the model developer, the end user of the model, and ``imposed users'' (individuals or groups who are affected by the model's decisions, outcomes, or recommendations). The model developer may consider each of the first three contexts~---~ Model Debugging and Validation; Model Selection; and Mental Model and Model Understanding~---~as they iterate in development. In contrast, end users will typically participate in Model Selection, Mental Model and Model Understanding, or Human Machine Teaming. Model Feedback, Challenging, and Prescription is of greatest interest to imposed users.  They may also be interested in the Mental Model and Model Understanding use case. Figure~\ref{fig:use_case_diagram} illustrates the overlap of relevant use cases across the three types of users. Our taxonomy is similar to that of Mohseni et al~\cite{mohseni2018survey}, although we distinguish imposed users from end users and do not distinguish end users from ``data experts''.

% \begin{center}
% \begin{table}[h]
% \begin{tabular}{|l|l|l|l|}
% \hline
%               & Developer & End User & Impacted \\ \hline
% Debug/Validate & X         &          &          \\ \hline
% Select         & X         & X        &         \\ \hline
% Understand     & X         & X        & X         \\ \hline
% Team           &           & X        &          \\ \hline
% Challenge      &           &          & X        \\ \hline
% \end{tabular}
% \end{table}
% \end{center}

In addition to considering the above users and use cases, i.e., the context in which an explanation is used, we also need to design controlled experiments with falsifiable hypotheses. We believe the ``gold standard'' XAI evaluation experiment should be one where all participants perform the same task, but a randomly assigned group of participants performs the task without the assistance of the explanation being evaluated. By comparing the performance of groups with and without the explanation, we can make claims about the benefit of that explanation for the task. The hypothesis (that the explanation is useful) is falsified when there is no significant difference between these groups.

Not all experiments in this research domain should follow this rigid experimental design, but we suggest following some basic guidelines. Researchers should first consider the three key components of their system: the human, the machine, and the explanation. Next, researchers should determine what can be measured (or is meaningful to measure) using different combinations of these components. Below are examples of four combinations of these components and how they provide us different information:
\begin{itemize}
    \item \ho: baseline performance of human at the task (the fully manual scenario)
    \item \mo: baseline performance of machine at the task (the fully automated scenario)
    \item \hm: baseline performance of the system when the user can rely on the machine learning output
    \item \hme: the performance of the system when the explanation and machine learning output are available to the user
\end{itemize}

Thus, one should compare the performance of \hme group against \hm  group. The \ho and \mo provide additional context for this comparison. For example, there may be a significant benefit of the explanation, but perhaps the \mo performance is greatest, indicating that the system should be fully automated. 

Of course, researchers should use their best judgement about what combinations are meaningful or practical for their specific applications. For example, Yang et al.~\cite{yang2020visual} designed an experiment in-line with this framework. However, the researchers decided not to measure the \ho condition because there was no reasonable expectation that unaided participants had the expertise to perform the task (identifying the species of a tree given a picture of a single leaf). The researchers decided to measure \hm performance as a baseline instead, which they found closely aligned with \mo performance, indicative of overtrust. In that study, the \hme performance was greater than both of the baselines, providing strong evidence of the benefit of the explanation.

The remainder of this section is organized around the use cases previously discussed and hypothetical downstream tasks for evaluation purposes. For each downstream task we describe a pseudo-experiment that is intended to provide inspiration for researchers who wish to conduct the scientifically rigorous  research we have argued for in our position statement.

\subsection{Model Debugging and Validation}

\usecase{Model Debugging and Validation} is a developer-focused use case that leverages explanations as a means of improving the model. Here, explanations are used to provide insight into the mechanisms of complex ``black box'' models, e.g., neural networks or other deep learning models \cite{castelvecchi_2016}, to identify flaws or biases in the algorithm or the training data that can be addressed in development. For example, a developer may use explanations of model decisions of varying confidence to probe whether the model relies strongly on non-actionable or biased features and identify constraints that are necessary to implement within the model (as such models should not be used in most, if not all, settings~\cite{ustun2019actionable}). 
    
\task{Given an incorrect model decision and corresponding explanation, determine the reason the model made a mistake.}%\\}
%\paragraph{\textbf{Task: }Given an incorrect model decision and corresponding explanation, can you determine the reason the model made a mistake?\\}
\taskdetails{A set of model mistakes are coded by the research team in order to establish a ground truth. In the case of a classification task, these coded mistakes may be an annotation of the image qualities that misled the model, e.g., occlusion of the subject, pixellation, and artifacts in the image as well as qualities of the model behavior, e.g., misplaced model attention, lack of training examples. Inter-rater reliability is established on the coded mistakes and the images are presented to users to determine if explanation methods are useful in helping users determining why a model made an error. Users' ability to correctly describe the reason for the model's mistakes are measured with and without the explanation.}

\task{%Given a model and explanation, i
Identify if the model will improve from additional training.}%\\}
\taskdetails{
% Deciding how much data to train on, or how much computational resources to use for training, is a common problem. Explanations may be helpful to determine if the model has learned useful and generalizable concepts.
Given three datasets, a machine learning model is trained on one and the user is presented with data in all three sets and the performance of the model, e.g. F-score on validation (from train) and the two test sets. The user is asked whether the performance on the test set would significantly improve if the model can include the second test set in its training.
Both train and test examples are available to the user, and in the \hme condition, the user can view explanations of model decisions on the original train and test data.
The ground truth is measurable because the difference in model performance on the smaller and larger training set are available. A variant of this experiment would ask the user to estimate the difference in performance. 
A challenge of this experimental design is the creation of datasets with suitable differences in performance.
}

\task{Given a model that exploits artifacts or loopholes of the data, describe the model's behavior.}%use explanations to describe the model's behavior.}
\taskdetails{Model ``intelligence'' has been reasonably challenged in recent years by the discovery of `Clever Hans' behaviors. This reveals the model's reliance on features of the data that humans would consider unintuitive (such as source tags in images) and are threats to generalizability \cite{Lapuschkin_2019}. In this task, users explore the model explanation to describe how decisions are being made about classes within the data. Performance is measured by determining whether users are able to discover undesirable behaviors in the model's decision-making, such as identification of trains by spotting rails, boats just by identifying water, or wolves by focusing on snow \cite{ribeiro2016should}.} A control condition, i.e. no explanation, is possible, but would require showing the user many correct and incorrect classifications to give the user an opportunity to understand the model based solely on behavior.

\subsection{Model Selection}
% need more concrete tasks here
% is A > B?
% rank A, B, C, ... N
% estimate precision/recall/f-score of A

Our second use case, \usecase{Model Selection}, is typically the focus of trust and explanation analyses when considered together because it seeks to answer the intuitive question of ``do I \textit{trust} this model enough to use it?'' or ``do I \textit{trust} this model more than another, and thus should use it instead?'' Although we argue this is a narrow application of trust and explanations together, it is a common (and important) use case to consider.
We note that the model selection use case is more complicated (and potentially problematic) when a second predictive or generative model is required to generate the explanations rather than using artifacts of or features extracted directly from the model under assessment (e.g. captions generated from image inputs and model decisions to explain model decisions).

\task{Determine which model is better suited for a given task.}%?\\}
%\paragraph{\textbf{Task: }Given two models, which is better?\\}
\taskdetails{In this task, users consider two models and either the decisions alone or the decisions alongside their accompanying explanations. Users' performance identifying which model performs better on an unseen test set across the two groups can be used to quantify the quality of the accompanying explanations --- e.g., by measuring whether using the explanations to identify whether the model decision was right (or wrong) for the right (or wrong) reasons enabled users to better distinguish which model is best suited for the task.

As a variant, users may be asked to determine which of the models would best extend to a specific out-of-domain task, e.g., classifying foods after seeing classification examples of animals or classifying posts on Twitter after seeing classification examples using Reddit data. %\notes{I think this can be interesting, but I am still working out how performance might be measured differently than the in-domain task.}
These tasks can intuitively be extended to an experiment ranking multiple models, all of which can be presented and paired with or without an explanation method and outcomes.
}
%Two models $A$ and $B$ are given the same input $x$, yielding outputs $y_A$ and $y_B$ which may not agree. For a set of inputs paired with explanations of the model decisions, users will decide which model is best suited for the task.

%
\subsection{Mental Model and Model Understanding}

The \usecase{Mental Model and Model Understanding} use case relates to whether a user builds an accurate mental model, that is, a mental model which functionally mirrors the overall behavior and decision making of the machine learning model. Visualizations are a popular type of explanation for building and eliciting mental models.
Mental models are important for developers to understand their own models. They are also critical for end users who use machine learning to gain insight about a new domain or to complete a task. Accurate mental models can also be helpful for imposed users who are affected by model decisions, for example, if they are denied services because of the model's classification of their profile or history. These users may want to understand how the model makes decisions in order to set expectations of how the model will impact the user. 

This use case differs from \textit{Model Selection} or \textit{Model Debugging and Validation} in that the focus shifts from explaining specific decisions to comprehending the relationships between the model input and output, or understanding the inner workings of a model's decision-making. This use case is challenging because there are no ways to directly observe a user's mental model, and appropriate metrics for measuring understanding are still widely debated \cite{UnderstandingToon}. 

% need more concrete tasks here
% is A > B?
% rank A, B, C, ... N
% estimate precision/recall/f-score of A

%* the user can see all of the original training data
%* the user can see all of the potential new training data
%* TASK: does performance (how defined?, discrete bins of model performance) increase when trained on new data?
%* user can see explanations of both new and old data (or neither for control)
%* ground truth: delta in model performance is measurable
%* challenge: creating situations where performance is or isn't increased

\task{Given examples of past behaviors, extrapolate what a model will do given unseen inputs.}
\taskdetails{
 In this task, users are given a series of inputs, and the corresponding model outputs. Then, users are presented with a series of previously unseen inputs.  A randomly assigned subset, representing the \hme condition, are given the corresponding explanations. The control group, representing the \hm condition, is not provided the explanations. Users are then asked to either select from a list, or describe in their own words, their expectation of model output. More accurate user predictions of the model output for the altered input provides evidence of the quality of the mental model built with the support of the explanation.
 
 Some variations of this task include using different types of input for the user to base their extrapolations on. For instance, the set of inputs may consist of all new inputs, none ever seen in the initial series of inputs with associated outputs, or a mix of inputs seen before and new inputs. Ribiero et al.~\cite{Ribeiro2018AnchorsHM} used all new inputs when testing a tool which presents a visual summary of why the model made a specific classification. In their user study, users were asked to predict a model's output first without an explanation, then presented with a set of decisions with explanation and finally asked to perform one more round of predicting the model's outputs. 
 
 Another variation could be in the temporal dimension of extrapolation. For a model such as an image classifier, any input and output can happen in any order. However, with a model such as a reinforcement learning agent, the task could consist of predicting the agent's immediate next move, or any number of time steps in the future, as explored by Anderson et al.~\cite{anderson2019explaining}. This type of variation would explore if a user's mental model is accurate enough to predict the model's future choices, and how far into the future that mental model accurately extends.  }

%Move this here as a variant instead of a separate task?
%As a variant, user groups are given separate explanations of the same model $A$ and can progress through as many training examples showing explanations for model decisions as they'd like before continuing to the experimental task. This task consists of selecting the expected model output $A(x)$ given input $x$. Feedback is given on whether the user chose the right output, and the user is free to move back to the training cycle at any time.}
 
\task{Given information about a model's past performance, match the model to a novel output.}%Given an explanation of a decision, can you tell which model it corresponds to}%?\\}
\taskdetails{When the context of decisions belonging only to a specified model or models is explicitly set, as in the above described task, users may overestimate how well they understand a model. In this task, user groups have that narrowed context removed, and are asked to differentiate between multiple models. User groups are given a series of input-output pairs for a set of models during a training phase. It is specified which model created each output to help users build mental models. After the training phase, users are presented with a new, previously unseen set of inputs and corresponding model outputs. Users are asked to match the models to the new decisions, or to identify if none of the previously presented models would have produced the given decision. This task is repeated including explanations for the decisions made by the model. 

The \hm condition would be represented by a user group receiving no additional explanation of the model's outputs during the process of building a mental model. The \hme condition would be represented by a group of users who get an explanation of each model's outputs in the mental model building phase. Any improvement in ability to correctly correspond models to the new decisions would signal that explanations help users to better understand the model. 

One variant of this experiment would be to test users' understanding of a single model by purposely altering the output of the model and testing if users can identify if the output is incorrect, and what part of the output is incorrect. Chang et al.~\cite{ChangReadingTeaLeaves2009} use this variation in a user study of a model which sorted documents into topics, which are defined by a set of words. Users were tested to see if they could detect manipulation by researchers of both the words describing a topic, and the topic assigned to a document. 

Another variant of this task would be to change the training process, so that instead of using the input-output pairs, the users would be given a global explanation or description of each model. For example, users may be told that an image classifier was trained on a specific data set of birds, and has been observed to rely heavily on beak shape and size in its classifications. }

\task{Given a model's past behavior, identify if explanations speed up user's creation of a mental model.}%\\}
%\paragraph{\textbf{Task: } How quickly/accurately can users  build mental models of model performance given different explanation types?\\}
\taskdetails{This task consists of selecting the expected model output given a previously unseen input (evaluation phase) after having studied examples of inputs and the corresponding models outputs (learning phase). Users in the \hme group are also provided an explanation corresponding to each model input-output pair. Users are informed that they will be timed and can proceed through as many sets of inputs and outputs as they want. Users can switch to the evaluation phase at anytime where real-time feedback is given on whether the user chose the right output. The user is free to move back to the learning phase to study more sets of known inputs and outputs before returning again to the evaluation phase. The total time spent, number of cycles, or accuracy in predicting the model behavior can be compared across groups to determine the benefit of the explanation. 
In a variation used by Lim et al.~\cite{LimWhyWhyNot2009}, a set number of examples are presented in a learning phase and users are timed according to how long they spend in the learning phase. Once users move on to the evaluation phase, they cannot return to the learning phase, and the time taken to answer each question in the task phase is also recorded. Lim et al. compared groups of users receiving explanations during the learning phase to groups getting no explanations. Alternatively, this task can be used to compare the efficacy of different explanations, which could be imperative in safety-critical environments.
}

% \paragraph{\textbf{Task: }\\}

\subsection{Human Machine Teaming}
\usecase{Human Machine Teaming} distinguishes models as teammates, beyond simple tools. Here, high-quality explanations can elevate models to act as teammates by providing more insightful and actionable recommendations. Essentially, good explanations can serve as the models response to ``explain your work'' or ``why?'' queries and assist users in complicated tasks or alleviate the cognitive load of human teammates. 
As human machine teaming necessarily generates more specific use cases than the tasks enumerated in previous sections, this section utilizes more specific use cases, that can of course be generalized to other domains.

\ignore{
%As introduced above, the baseline case of {\it Human + Machine} should be compared to 
%the efficacy of explanations can be the diff
Delineate between teammate or tool 
-- on the job or before the job 
-- who is using the model?
-- alleviate cognitive load of user?
-- level of automation of task
}

\task{Identify whether users should accept or reject model decisions.} % from a given model.}%Given an input, explanation, and a model decision, identify whether to accept or reject the decision.}%\\} 
\taskdetails{Given a series of inputs, users are tasked with agreeing or disagreeing with a model's output, accompanied by an explanation of interest for a subset of user groups. Performance in this case is measured based on the accuracy of final decisions selected by the user, which is either the model decision when users agree with the decision or the user-selected decision when users disagree with the model. This task effectively calibrates the users' \textit{appropriate trust} in the model. Performance of the \hme group is compared directly to the \hm group. This experiment requires that the \ho performance be less than the \mo performance to show a benefit, such as classifying tree leaves in images~\cite{yang2020visual} (instead of everyday objects) or performing quality control in an assembly line scenario~\cite{bansal2019beyond}.

An extension to this task can consider reduction of the cognitive load for users as the desired outcome and user performance can also be measured using proxies for cognitive load such as the number of correct judgements made by the user in a limited time period or the time needed to complete a specified number of correct judgements. 

%\notes{* same task, just change the context?}

}

\ignore{
\task{Determine whether the cognitive load for the user is reduced if explanations are provided alongside model decisions, recommendations, or other output.}%Given an input, a model decision, and an explanation, determine whether the cognitive load for the user is reduced.}%\\}
%\paragraph{\textbf{Task: }Given an input, a model decision, and an explanation, is the cognitive load for the user reduced?\\}
\taskdetails{As an extension to the previous task, users are given a series of model decisions and asked to accept or reject the recommendations of the model and researchers measure cognitive load of the user while completing the tasks. In contrast to the previous task, performance is measured using a proxy for cognitive load for the user, e.g., the number of correct judgements made by the user in a limited time period or the time needed to complete a specified number of correct judgements.}
}

\task{Determine whether properly abstracted explanations improve human experience and performance in an autonomous driving scenario.}
\taskdetails{One of the challenges when designing model explanations lies in understanding which end user the explanation is being designed for. For example, levels of abstraction change drastically if explanations are designed to target the software engineer responsible for autonomous navigation and collision avoidance rather than the driver sitting behind the wheel. In this task, user groups \textit{driving} simulated autonomous vehicles would be provided with simplistic explanations of the car's behavior as a driving test takes place such as warnings about poor object detection in fog or reduced traction in sharp curves. These explanations would be given when environmental dangers are encountered during the simulation. A control group would receive no explanations. Performance can then be measured based on the users' situational awareness (SAGAT, etc. \cite{endsley_1995}), attention switching from the road to the explanations, trust questionnaires, and other physiological indicators (e.g., heart rate, eye motion).

An extension to this task can present the explanations of the car's behavior in a ``user manual'' style before the driving test begins and measure how often the drivers accurately respond to hazardous conditions with no real-time input.}

\task{Determine whether explanations increase the efficiency of a human machine team.}
\taskdetails{
Healthcare has been a focus of recent work in human machine teaming, and serves as an exemplary application domain. Studies have shown the effectiveness of explanations on both trust and interpretability in ML models focused on medical diagnosis~\cite{Diprose2019}. As the availability and complexity of medical technology increases, physicians may find themselves in need of machine agents who can help them narrow in on useful treatments and diagnoses. In this task, the user works with an automated physician's assistant who makes recommendations for data collection, testing, and diagnoses during a physician-patient interaction. 

The model presents choices to the physician with accompanying explanations of these choices in the \hme case, such as visuals of specific patient data and its risk contribution for certain diagnoses. This can present the issue of branching, where presented choices may generate a longer path to the end goal with the potential of detours that do not offer viable paths to the solution. Two control groups exist for this task, that being \hm groups with no generated explanations of proffered choices and \ho groups which receive no model assistance. How long it takes user groups to arrive at the correct diagnosis, the cost of that treatment, and how many incorrect branches were explored are all viable performance metrics to establish whether the explanations of model benefited users. As this task requires trained experts, i.e. physicians, the \ho baseline is meaningful.

A wide array of extended tasks can be generated from this initial example. Time limits can be imposed on the entire task to consider cognitive load on the users, and to examine whether or not user groups are able to finish the task at all. The use case can also be extended into the work allocation domain, wherein the model recommends actions and gives explanations (or does not) that have cascading consequences on the subsequent work tasks, creating a dynamic environment that is shaped by the human-machine team and which can end up in any number of end states with measurable utilities. Additional physiological measures, post-experiment questionnaires, and human factors metrics (e.g. situational awareness or perceived cognitive load) can then be applied to understand the usefulness of these model explanations along different dimensions. }

% * measure classical things (precision, recall, etc) - did we get all of the documents the user wants? did we retrieve all information the user needed? * using a model, can humans get from A to Z more quickly than a human alone in a multi-step task? (e.g. "Wikipedia Race" game where goal is to get from source to target wikipedia page in the fewest links)* chat bot: you have to find out some Q&A. If the chat bot only gives a response to the question, or it gives the response AND an explanation * 

% * branching - if there are always multiple options = potential for rabbit holes, or user has to choose best -- do you get to the right solution quickly or all? do you finish within the time limit or not?
% * diagnosis -- common cold vs. cancer? -- automated physician's assistant, iterative process -- run multiple tests to identify correct diagnosis. measure - how long it takes to complete, do your reach the correct answer, how many incorrect classifications, how many unnecessary tests are completed, accumulate time/\$ costs?

% \task{Determine whether human team leaders are able to more effectively complete work allocation with an explanation of model behavior.}
% \taskdetails{* model/teammate failure inputs: model explanation of the failure and the decision that caused it * A to }
% * stop task if there's task failures; mismatches between teammate and user goals; reallocation of task

\subsection{Model Feedback, Challenging, and Prescription}
% (trust in the system)

The \usecase{Model Feedback, Challenging, and Prescription} use case effectively considers \textit{trust in the system}, i.e., the model in context of the \textit{impact} on imposed users of its decisions and subsequent recommendations or actions taken. The need for effective explanation of model decisions for recourse is a natural response to the continued widespread application of artificial intelligence or machine learning models to supplement or automate tasks in domains where incorrect or biased recommendations can have significant human impacts. These domains include predictive policing~\cite{ensign2018runaway}, recidivism prediction~\cite{chouldechova2017fair,dressel2018accuracy}, and hate speech and abusive language identification online~\cite{park2018reducing,davidson2019racial,sap2019risk}. As an example of the recognized necessity of clear explanations for this use case, the European Union's GDPR directly addresses the ``right of citizens to receive an explanation for algorithmic decisions''~\cite{goodman2017european}.
%  The European Union's GDPR policy on "the right of citizens to receive an explanation for algorithmic decisions"
 %- predictive policing, recidivism bias, GDPR right to explanation
 %European Union’s new General Data Protection Regulation - "The GDPR’s policy on the right of citizens to receive an explanation for algorithmic decisions highlights the pressing importance of human interpretability in algorithm design"

\task{Given a model's decision, identify how to get a better outcome.}
\taskdetails{In this task, a user may be given an input and a decision from a model and asked to identify what has to be changed or updated in the input to get a better decision. User groups are either given an accompanying explanation for the decision of the unaltered input as well as or only the model output. This task touches on counterfactual explanations and the prescriptive use of algorithmic decision making systems. The ``What if tool''~\cite{wexler2019if} supports this type of counterfactual reasoning. However, the tool is oriented towards end users and developers~---~there is still a research opportunity to design explanations that support imposed users for this use case.

% TODO abrupt
Given the model is \textit{trustworthy}, providing the right outcome given the right data, but the user desires a different outcome, does the model explanation provide enough information for a user to identify the personal changes needed to obtain the desired outcome? \textit{Glass-Box}~\cite{sokol2018glass} is an example of such a downstream task in practice -- users, given a pre-established persona, probe a loan application system that provides contrastive, counterfactual explanations to understand and challenge the model's automated decisions.
}

% Feedback - the model had wrong or incomplete data so wrong outcome
% Challenging - it had the right data, bad/biased inference so wrong outcome
% ?? - it had the right data, made the right outcome, I want to change it 

\task{Given a model's decision, determine if the decision was based on incorrect data, biased data, or bad inference.}
%\paragraph{\textbf{Task: }Given a model's decision about me, can I tell if the decision was based on incorrect data about me, biased data, or bad inference?}
\taskdetails{User groups are provided an overview of the training examples for a given model and a series of input-output pairs where the model output was incorrect. Users are then asked to identify \textit{why} the model provided an incorrect response or what methods might be employed to correct the decision (e.g., more training examples for edge-case inputs, removal of misleading or biased training examples, or removal or preprocessing of flawed inputs) and whether the change required is a reasonable expectation (reduce the level of outstanding credit to receive a new load) or an indication of bias (change your gender/race to receive a new loan). A control group (\hm) will not receive model explanations, but would have access to the training dataset in order to contrast performance with the \hme group to examine the benefits of the explanations. 
More accurate user predictions of the model output for the altered input provides a quantitative measure of the quality of the explanation to identify biases in the model -- whether the model is \textit{fair} across the affected population -- and also as a means to identify challenge-worthy individual decisions. 

% rudin2age
%
%  sokol2018glass
%  where users impersonate one of 10 loan applicants (to avoid a lengthy process of submitting personal details) and are able to interrogate and challenge an automated decision
}
% loan application denied - race/gender or outstanding credit?
% recommender systems - recommended movie (watched one thing and get biased recommendations through the rest --- explanation tags specific movie and you can provide feedback to not consider that)
% sometimes recommendations make sense but other times it doesn't -- watch the notebook because you watched the matrix (because both are popular movies)
% output = "we recommend X because you watched Y" 
% measure success based on whether the feedback improved the recommendation -- can we steer the model in the right direction? Does feedback improve the model? 
% Give users persona to follow and description of recent history

\section{Conclusions}
In this paper we have argued that trust in a machine learning model is a benefit of a useful and reliable system that employs that model. However, trust develops slowly over time, and to rely on trust as a metric for evaluating the value of an explanation is problematic and could lead to artificially inflated levels of trust to the users' detriment. We believe trust should only be measured in a longitudinal and empirical study considering the full system.

Instead, researchers should design for and measure utility. Utility-oriented evaluation encourages researchers to consider the broader context of the explanation, i.e., how it is intended to be used. It also encourages researchers to employ scientific methodologies to evaluate explanations, leveraging falsifiable hypotheses and objectively measurable quantities as evidence. Towards this end, we have suggested many pseudo-experimental designs involving ``downstream tasks'' that can be used to evaluate explanations in this manner. We hope the impact of this work will be to inspire many new experiments that solidify the scientific foundation relating humans, machines, and explanations.

\acknowledgments{
The authors wish to thank Matthew Taylor for helpful discussions around downstream tasks in reinforcement learning.}

\bibliographystyle{abbrv-doi}

\bibliography{main}
\end{document}